\documentclass[useAMS,usenatbib]{mn2e}
\usepackage{graphicx}
\usepackage{amssymb}
\usepackage[T1]{fontenc}
\usepackage{aecompl}

\title[Attenuation of super-soft X-ray sources by circumstellar material]{Attenuation of super-soft X-ray sources by circumstellar material}
\author[M.T.B. Nielsen \& M. Gilfanov]{M.T.B. Nielsen$^{1}$\thanks{E-mail:
mede@mpa-garching.mpg.de}, M. Gilfanov$^{1,2}$ \\
$^{1}$Max-Planck Institut f\"ur Astrophysik, Karl-Schwarzschild-Str. 1, Postfach 1317, D-85741 Garching, Germany\\
$^{2}$Space Research Institute of Russian Academy of Sciences, Profsoyuznaya 84/32,117997 Moscow, Russia}
\begin{document}

\date{Accepted -. Received \today; in original form -}

\pagerange{\pageref{firstpage}--\pageref{lastpage}} \pubyear{2015}

\maketitle

\label{firstpage}

\begin{abstract}
Recent studies have suggested the possibility of significantly obscuring super-soft X-ray sources in relatively modest amounts of local matter lost from the binaries themselves. If correct, then this would have explained the paucity of observed super-soft X-ray sources and would have significance for the search for single-degenerate type Ia supernova progenitors. We point out that earlier studies of circumbinary obscuration ignored photo-ionisations of the gas  by the emission from the super-soft X-ray source. We revisit the problem  using a full, self-consistent calculation of the ionisation state of the circumbinary  material photo-ionised by the radiation of the central source. Our results show that the circumstellar mass-loss rates required for obcuration of super-soft X-ray sources is about an order of magnitude larger than those reported in earlier studies, for comparable model parameters. While this does not entrirely rule out the possibility of circumstellar material obscuring super-soft X-ray sources, it makes it unlikely that this effect alone can account for the majority of the missing super-soft X-ray sources. 
We discuss the observational appearance of hypothetical obscured nuclear burning white dwarfs and show that they have signatures making them distinct from photo-ionised nebulae  around super-soft X-ray sources imbedded in the low density ISM.
\end{abstract}


%
\section{Introduction} \label{Sect:Introduction}
The topic of a massive white dwarf accreting hydrogen-rich material from a non-degenerate companion and processing that material into heavier elements on its surface is of relevance to the understanding of a number of astrophysical objects, including symbiotics (e.g. \citealt{Mikolajewska.2012}) and single-degenerate type Ia supernova progenitors (e.g. \citealt{Maoz.et.al.2014}). In the the case of the single-degenerate scenario, the accretion process is necessary for the white dwarf to grow in mass until it becomes massive enough to undergo thermonuclear runaway in its degenerate core and explode. The steady burning of accreted hydrogen-rich material on the surface of a massive white dwarf is believed to be possible only in a narrow interval of mass-accretion rates (\citealt{Nomoto.1982,Shen.Bildsten.2007}), and the process is expected to emit copious amounts of super-soft X-rays \citep{van.den.Heuvel.et.al.1992,Kahabka.van.den.Heuvel.1997}. However, several studies have established a discrepancy of one to two orders of magnitude between the expected and observed number of systems, as well as the integrated luminosities of super-soft X-ray sources in old galactic populations, provided type Ia supernova progenitors are in fact single-degenerates \citep{Gilfanov.Bogdan.2010, Di.Stefano.2010}.

The missing super-soft X-ray sources could perhaps be explained if the X-ray emitting systems were obscured by local material. There is growing evidence for the presence of circumstellar material around at least a subset of type Ia supernova, as inferred from supernova spectra (\citealt{Gerardy.et.al.2004,Patat.et.al.2007,Simon.et.al.2009,Sternberg.et.al.2011,Cao.et.al.2015}), as well as from historical supernova remnants \citep{Borkowski.et.al.2006,Chiotellis.et.al.2012,Broersen.et.al.2014}

Circumstellar material around type Ia supernova progenitors could be supplied in a variety of ways, e.g. by a wind from either of the binary components, stellar pulsations of the companion, incomplete accretion of wind-Roche lobe overflow \citep{Mohamed.Podsiadlowski.2007} or tidal interactions between the components \citep{Chen.et.al.2011}. The effect of circumstellar obscuration on the observability of super-soft X-ray emission from progenitor systems was investigated by \citet{Nielsen.et.al.2013,Wheeler.Pooley.2013}. Both studies suggested that relatively modest amounts of circumstellar material could obscure super-soft X-ray sources to the point where they would no longer be detectable with current X-ray telescopes. However, both studies either ignored or used rather simple assumptions concerning the effect of photo-ionisation of the circumstellar material, and specifically did not take photo-ionisations of metals into account. The present study seeks to remedy those shortcomings by conducting a full treatment of the photo-ionisation of the obscuring circumstellar material, to get a more accurate handle on how much circumstellar material is required for obscuration.

The reader should note the terminology used in the following: \textit{mass-transfer rate} denotes the rate of mass transferred from the donor to the accretor, regardless of whether the transferred material is retained on the accretor, whereas \textit{mass-accretion rate} -- or just accretion rate -- is the rate of material transferred \textit{and} retained on the accretor; lastly, \textit{mass-loss rate} refers to the rate of mass lost from the binary into the circumbinary region where it may obscure the super-soft X-ray source. We use the symbol $\dot{M}_w$ for the latter.

In Section \ref{Sect:Model} we describe our model, Section \ref{Sect:Results} presents the results of our calculations, while Section \ref{Sect:Discussion} discusses the implications of the results. Section \ref{Sect:Conclusion} concludes.

%
\section{Model} \label{Sect:Model}
The geometry of the model examined in this study is identical to the one considered in \citet{Nielsen.et.al.2013}. The primary difference between that study and the present one is the incorporation of full photo-ionisation calculations. In particular, we consider  a super-soft X-ray source inside a spherically symmetric shell of material, surrounded by vacuum. Throughout the following, we assume that the source is a blackbody, with effective temperature $kT_{\mathrm{eff}}$ between 30 eV and 150 eV, and luminosities $L_{\mathrm{bol}}$ between $10^{37}$ erg/s and $10^{38}$ erg/s, parameters characteristic of 'canonical', persistent super-soft X-ray sources \citep{van.den.Heuvel.et.al.1992,Kahabka.van.den.Heuvel.1997}.

The gas shell surrounding the source is defined by an inner and outer boundary ($r_{\mathrm{inn}}$ and $r_{\mathrm{out}}$, respectively), and the density is given by an $r^{-2}$-law, characteristic of spherically symmetric, constant-velocity wind mass loss:
\begin{eqnarray}
 \rho(r) = \frac{\dot{M}_w}{4 \pi r^2 u_w}, \phantom{........} r_{\mathrm{inn}} < r < r_{\mathrm{out}}    \label{Eq:density.structure.from.wind}
\end{eqnarray}
where $\rho(r)$ is the mass density as a function of radius, $\dot{M}_w$ is the mass-loss rate into the circumstellar region and $u_w$ is the velocity of the wind, assumed to be constant in the region of interest. Inside the inner and outside the outer boundaries we assume vacuum. The inner radius can be envisioned as equivalent to the orbital radius of the binary (see also Sections \ref{Sect:Results} \& \ref{Subsect:Caveats}). Realistic density profiles are likely to be more complex than this, and clearly depend on the mass-loss history of the binary, however, the simple $r^{-2}$ profile is a useful first approximation and also facilitates comparison with previous results.

\subsection{Order-of-magnitude estimates}
The total mass of the gas shell is 
\begin{eqnarray}
M_w &=& \frac{\dot{M}_w}{u_w}\ \left( r_{\mathrm{out}}-r_{\mathrm{inn}}\right)\\
\nonumber
&=& 9.5\times 10^{25} \left(1-\frac{r_{\mathrm{inn}}}{r_{\mathrm{out}}}\right)  \dot{M}_{w,-6}\ u^{-1}_{100}\ r_{\mathrm{out,AU}} \ {\mathrm{g}}
\end{eqnarray}
where $\dot{M}_{w,-6}$ is the mass loss rate in units of $10^{-6}$ M$_\odot \mathrm{yr}^{-1}$, $u_{100}$ is the outflow velocity in units of 100 km/s and  $r_{\mathrm{out,AU}}$ is the outer radius of the shell in astronomical units.

The column density of the gas shell is:
\begin{eqnarray}\label{Eq:column.from.rinn.to.rout}
N_{\mathrm{H}}  &  = & \frac{\dot{M}_w}{4 \pi u_w \mu m_p} \left( \frac{1}{r_{\mathrm{inn}}} - \frac{1}{r_{\mathrm{out}}} \right)\\
\nonumber
&  = & 1.4\times 10^{22}\left(1-\frac{r_{\mathrm{inn}}}{r_{\mathrm{out}}}\right)  \dot{M}_{w,-6}\ u^{-1}_{100}\ r^{-1}_{\mathrm{in,AU}} \ {\mathrm{cm}}^{-2}
\end{eqnarray}
where $\mu\approx 1.4$ for Solar abundance, and $m_p$ is the mass of the proton. A column density of $\sim 10^{22}$ cm$^{-2}$ of neutral material of Solar abundance can easily obscure emission from a super-soft source with temperature of $\sim 100$ eV, as discussed in \cite{Nielsen.et.al.2013,Wheeler.Pooley.2013}. However, photo-ionisation of the wind material by the emission from the super-soft source can significantly reduce its opacity. 

Similar to the classical Str\"omgren sphere problem, the effect of the photo-ionisation can be estimated, to the first approximation, by comparing the total rates of photo-ionisation and recombination in the gas shell:
\begin{eqnarray}
& & \dot{N}_i = \int_{h\nu_0} L_\nu d\nu=f\frac{L}{h\nu_0} 
\label{eq:rate_i}
\\
& & \dot{N}_r =  \int n^2\alpha_{rec}\  d^3r = \frac {\dot{M}^2_w\ \alpha_{rec}} {4 \pi u^2_w\mu^2 m_p^2 }
\left( \frac{1}{r_{\mathrm{inn}}} - \frac{1}{r_{\mathrm{out}}} \right)
\label{eq:rate_r}
\end{eqnarray}
where $h \nu _0$ is the photo-ionisation threshold of the main absorbing ion, $\alpha_{rec}$ its recombination coefficient, $L$ is the luminosity of the central source and the factor $f$ depends on its radiation spectrum. In Eq. (\ref{eq:rate_i}), it was assumed that the gas shell  is optically thick to the radiation of the central source (in the absence of photo-ionisation).  If the photo-ionisation rate $\dot{N}_i$ exceeds the recombination rate  $\dot{N}_r $, the shell material will be fully ionised and  therefore transparent to the radiation of the central source. The critical value of the mass outflow rate is  determined by the condition $\dot{N}_i=\dot{N}_r$. At  lower $\dot{M}_w$ the gas shell is fully ionised and transparent, whereas at higher outflow rates it is opaque to the radiation of the central source. For monoatomic hydrogen gas, the  value of the critical outflow rate can be computed straighforwardly; with $h\nu_0=13.6$ eV and $\alpha_{rec}=2.6\times 10^{-13}$ cm$^3$sec$^{-1}$ (case B) one obtains:
\begin{eqnarray}\label{eq:mdot_crit_H}
 \dot{M}_{w,\mathrm{crit}} = 2.2\times 10^{-5} (fL_{38})^{1/2} u_{100}\ r^{1/2}_{\mathrm{in, AU}} \nonumber\\
 \times \left( 1-\frac{r_{\mathrm{inn}}}{r_{\mathrm{out}}} \right)^{-1/2}  {\mathrm{ ~ M_\odot/yr}}
\end{eqnarray}
where $L_{38}$ is the luminosity in units of $10^{38}$ erg/s. For the central source temperature of $kT_{\mathrm{eff}}=100$ eV, for example, $f\approx 5\times 10^{-2}$ and the leading factor in Eq. (\ref{eq:mdot_crit_H}) becomes $4.9\times 10^{-6}$ M$_\odot \mathrm{yr}^{-1}$. Eq. (\ref{eq:mdot_crit_H}) is sufficiently accurate for hydrogen gas shell, but gives only an order of magnitude estimate for gas of solar abundance. The latter case requires a more careful account of the ionisation state of the main absorbing species. However, Eq. (\ref{eq:mdot_crit_H}) predicts a nearly correct scaling of the critical mass loss rate (as confirmed by  Cloudy calculations in the following sections): 
\begin{eqnarray}\label{eq:mdot_crit}
 \dot{M}_{w,\mathrm{crit}} \propto L^{1/2} u_w\ r^{1/2}_{\mathrm{inn}} \left( 1-\frac{r_{\mathrm{inn}}}{r_{\mathrm{out}}} \right)^{-1/2}
 \end{eqnarray}
As one can see from Eq. (\ref{eq:mdot_crit}), the critical mass loss rate relatively weakly depends on the luminosity of the central source and the inner radius of the shell and is nearly  insensitive to the outer radius of the shell (for sufficiently large $r_{\mathrm{out}}/r_{\mathrm{inn}}$). The latter should be expected, given the steepness of the assumed density profile. 

The above considerations strictly apply to a stationary gas shell. In the case of the wind, (presumably) neutral material is continuously added at the inner edge of the shell. It is promptly\footnote{the photo-ionisation time at the inner edge of the gas shell is of the order $\tau_{ph}\sim 10^{-3}-10^{-2}$ s.} photo-ionised by the radiation from the central source. However, if the new material is added at a sufficiently large rate -- i.e. one comparable to the photon injection rate -- it will screen the rest of the shell from the ionising radiation, thus reducing its degree of ionisation and increasing its opacity. The importance of advection is determined by the ratio of the rate at which new material is added to the inner edge of the shell to the injection rate of  ionising photons:
\begin{eqnarray}
\frac{\dot{N}_{gas}}{\dot{N}_{phot}}=6\times 10^{-5} \dot{M}_{w,-6} \left( f_{-1}L_{38} \right)^{-1}
\label{eq:adv}
\end{eqnarray}
where $f_{-1}$ is the factor $f$ in units of $10^{-1}$. The ratio given in Eq. (\ref{eq:adv}) becomes of the order of unity at $\dot{M}_w\sim 10^{-2}$ M$_\odot \mathrm{yr}^{-1}$, which is much larger than the mass outflow rates considered here.

\begin{figure*}
    \centerline{
    \includegraphics[width=1.0\linewidth]{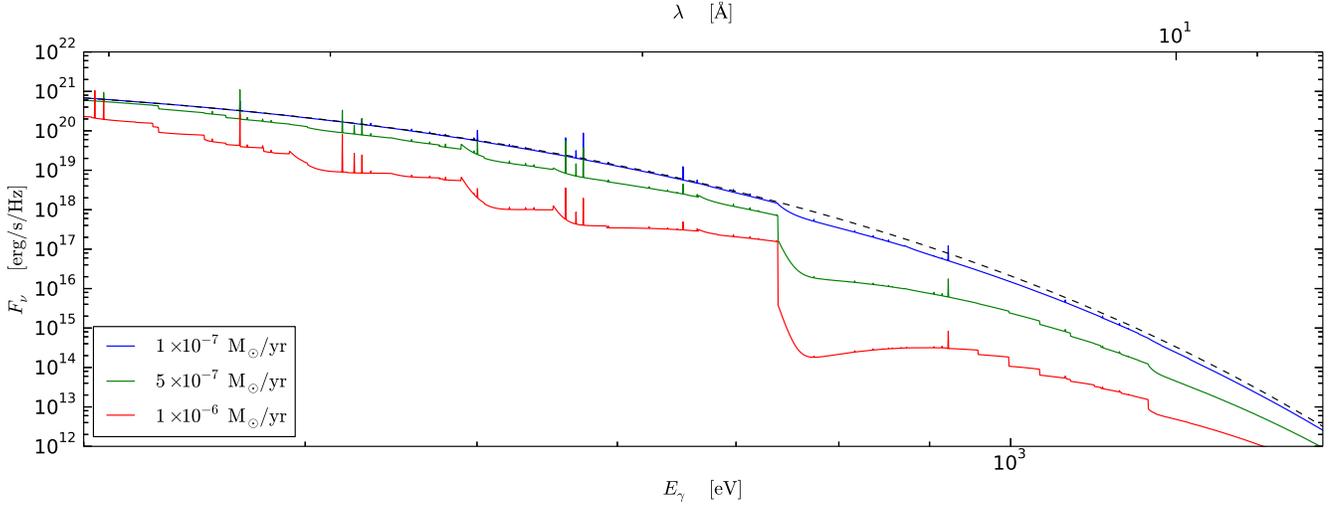}
    }    
    \caption{Attenuated spectra of the super-soft X-ray source (blackbody, $kT_{\mathrm{eff}}=50$ eV, $L_{\mathrm{bol}}=10^{38}$ erg/s) in the X-ray band for three values of the mass-loss rate. Parameters of the gas shell are $r_{\mathrm{inn}}=1$ AU, $r_{\mathrm{out}}=10$ AU, $u_w=50$ km/s. The dashed line shows the unabsorbed spectrum. Compare with the '1.0-10 AU' curve in the second panel from the top, middle plot in Figure \ref{Fig:atten_comb}.}
\label{Fig:absspectr_combined_1-10AU}
\end{figure*}

\subsection{Cloudy calculations}
For detailed photo-ionisation calculations, we used the one-dimensional photo-ionisation code Cloudy, version 13.02 (for a detailed description of the latest release of Cloudy see \citealt{Ferland.et.al.2013}). Cloudy calculates the ionisation and thermal state of a cloud of gas that is irradiated by a central source, by simultaneously solving equations of photo-ionisation and thermal equilibrium. It is assumed that atomic processes in the gas have become time-steady. This assumption is reasonable, at least in the context of  single-degenerate type Ia supernova progenitors, since all  microphysical time scales in the shell are much shorter than  time scales relevant to super-soft sources. In our simulations, we fix the density distribution  and compute the gas temperature self-consistently, as maintained by the photo-ionisation of the gas by the central source. For the gas density distribution, we use the built-in \texttt{dlaw wind} option in Cloudy which reproduces the density structure in Eq. (\ref{Eq:density.structure.from.wind}), and the {\tt sphere} keyword as appropriate for a closed geometry. The advection effects are not included in our calculations, as justified above. Cloudy's output files specify the emergent spectrum, temperature, pressure, and ionisation structure, in addition to a host of other data.

We run Cloudy for three values of $u_w$: 10, 50, 100 km/s. This spans the likely wind velocities expected from red giant and asymptotic giant star donors \citep{Owocki.2013}. For the purposes of this study we do not specify the nature of the wind. For each combination of $L_{\mathrm{bol}}$, $kT_{\mathrm{eff}}$, $r_{\mathrm{inn}}$, $r_{\mathrm{out}}$ and $u_w$ we run Cloudy for a range of values of $\dot{M}_w$ from $10^{-10}$ M$_{\odot}\mathrm{yr}^{-1}$ to $10^{-5}$ M$_{\odot}\mathrm{yr}^{-1}$.

The chemical composition of the material in the shell is Solar, as given in \citet{Grevesse.et.al.2010}. In Subsection \ref{Subsect:Dependence.on.chemical.abundances} we compare this with the abundances used in \citet{Nielsen.et.al.2013}.

For hydrogen and helium, Cloudy can be expected to give reliable results up to $n \sim 10^{19}$ cm$^{-3}$. However, for number densities $\gg 10^{10}$ cm$^{-3}$ the full treatment of a plasma containing heavier elements should not be considered reliable, see the Cloudy manual 'Hazy'\footnote{Sect.3.8 on p.35 in 'Hazy' Part 2 for Cloudy c13.1. The two Hazy manuals are downloadable with the Cloudy installation files on http://nublado.org/wiki/DownloadLinks.}. For comparison, the largest value of $\dot{M}_w$ for which one of our plots have data points ($10^{-5}$ M$_{\odot}$/yr at $r_{\mathrm{inn}}=1$ AU and $u_w=50$ km/s on Figure \ref{Fig:emission_line_luminosities_vs_Mdot}) corresponds to the number density of $1.9\times 10^{10}$ cm$^{-3}$, i.e. within the nominal density range of Cloudy.

%
\section{Results} \label{Sect:Results}
Cloudy's output can be used to reconstruct the spectrum of the system consisting of the super-soft X-ray source and circumstellar cloud, along with the ionisation structure of the circumstellar material. In the following, we use the term \textit{output spectrum} to mean the sum of the spectrum of transmitted radiation and the emission from the irradiated cloud. For the output data, we concentrate on the photon energy interval between 0.3 and 1.5 keV. Below 0.3 keV, even a modest ISM column effectively blocks any super-soft emission, and above 1.5 keV the emitted energy of a super-soft X-ray source is negligible.

To quantify how the attenuation depends on the various input parameters, we calculated the integrated emitted super-soft luminosity $L$ of the output spectrum for each input configuration and compared this with the integrated emitted luminosity of the unabsorbed super-soft source spectrum, $L_0$ (using the photon energy interval between 0.3 and 1.5 keV as integration limits). This yielded a transmission ratio $L/L_0$ between 1 (no obscuration) and 0 (complete obscuration) for each input configuration.

\subsection{Attenuation of super-soft emission} \label{Subsect:Attenuation.of.super-soft.emission}
To demonstrate the general trends in our results, Figure \ref{Fig:absspectr_combined_1-10AU} shows a comparison between the unabsorbed input spectrum of a $kT_{\mathrm{eff}}=50$ eV, $L_{\mathrm{bol}}=10^{38}$ erg/s blackbody source, and the output spectra in a cloud with inner and outer radii of 1 and 10 AU, respectively, and $u_w =$ 50 km/s, for three values of $\dot{M}_w$ from $1\times10^{-7}$ M$_{\odot}\mathrm{yr}^{-1}$ to $1\times10^{-6}$ M$_{\odot}\mathrm{yr}^{-1}$. For the smallest mass-loss rate shown the output spectrum is only weakly  attenuated. For larger values of $\dot{M}_w$, absorption edges of various neutral and ionised species appear in the spectrum along with their line and recombination emission. Among the absorption edges, the  most significant is {\sc Ovii} around $\sim$739 eV. At $\dot{M}=1\times10^{-6}$ M$_{\odot}\mathrm{yr}^{-1}$ the 0.3--1.5 keV flux is attenuated by a factor of $\approx 0.30$.

\begin{figure*}
    \centerline{
    \includegraphics[width=0.34\linewidth]{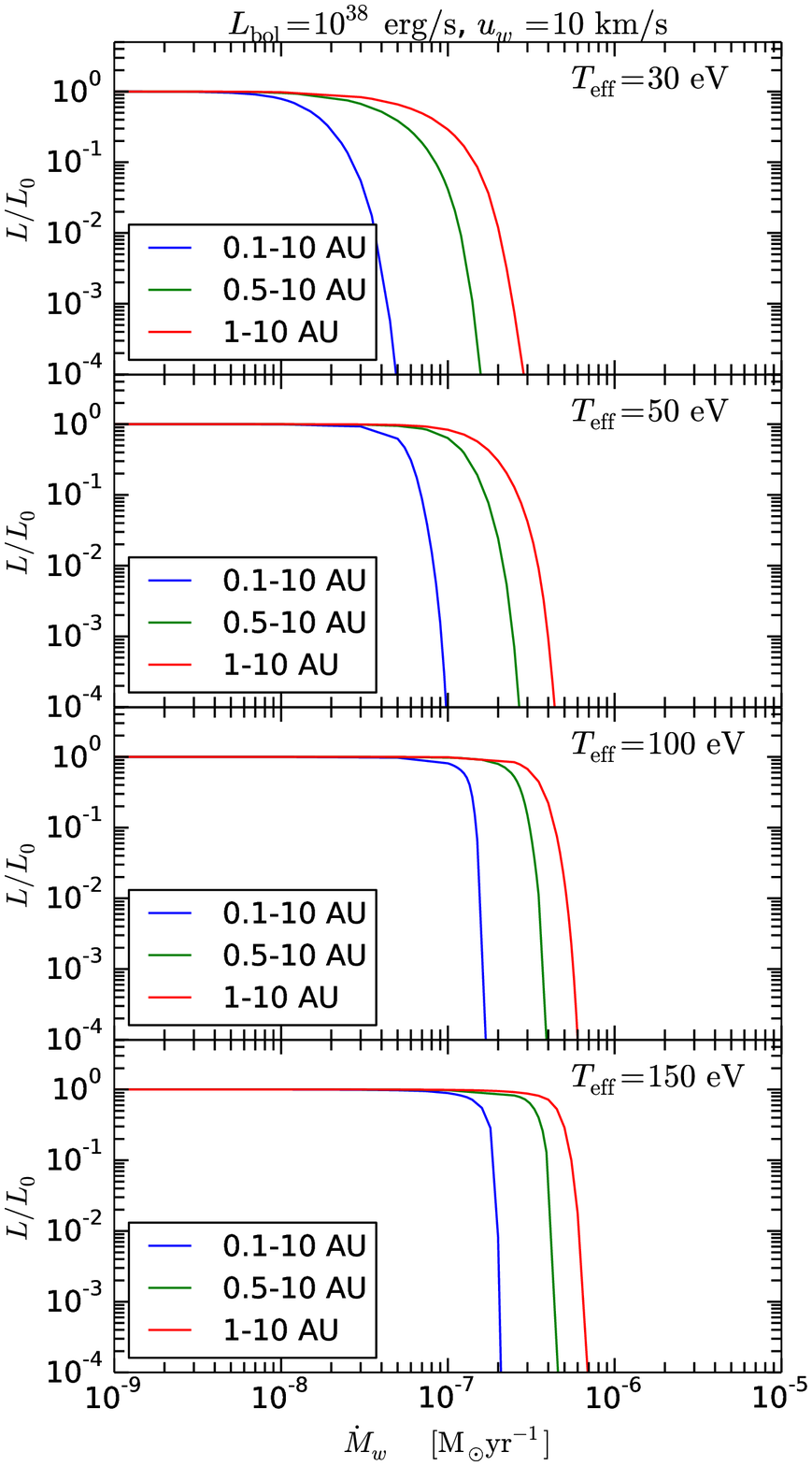}
    \includegraphics[width=0.34\linewidth]{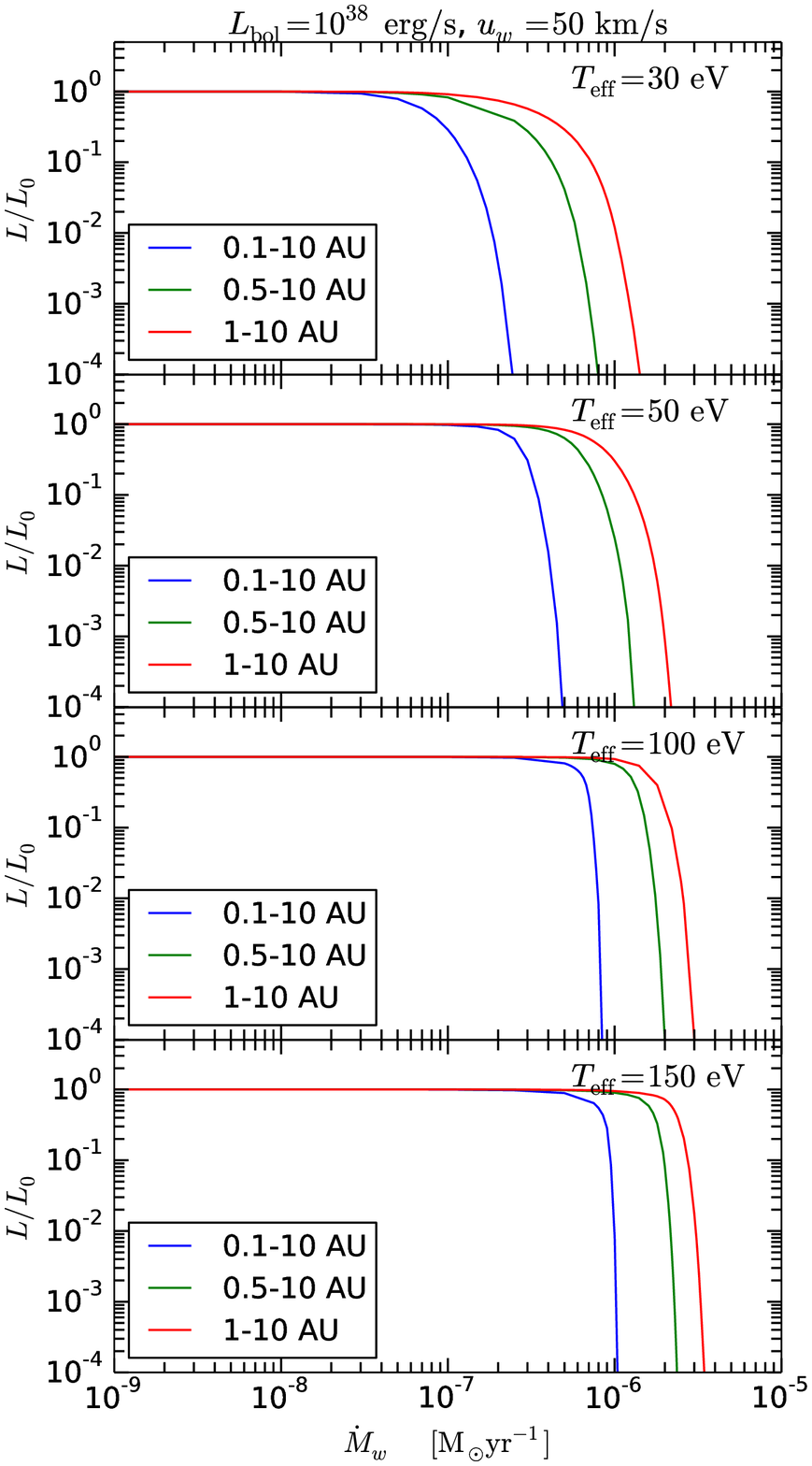}
    \includegraphics[width=0.34\linewidth]{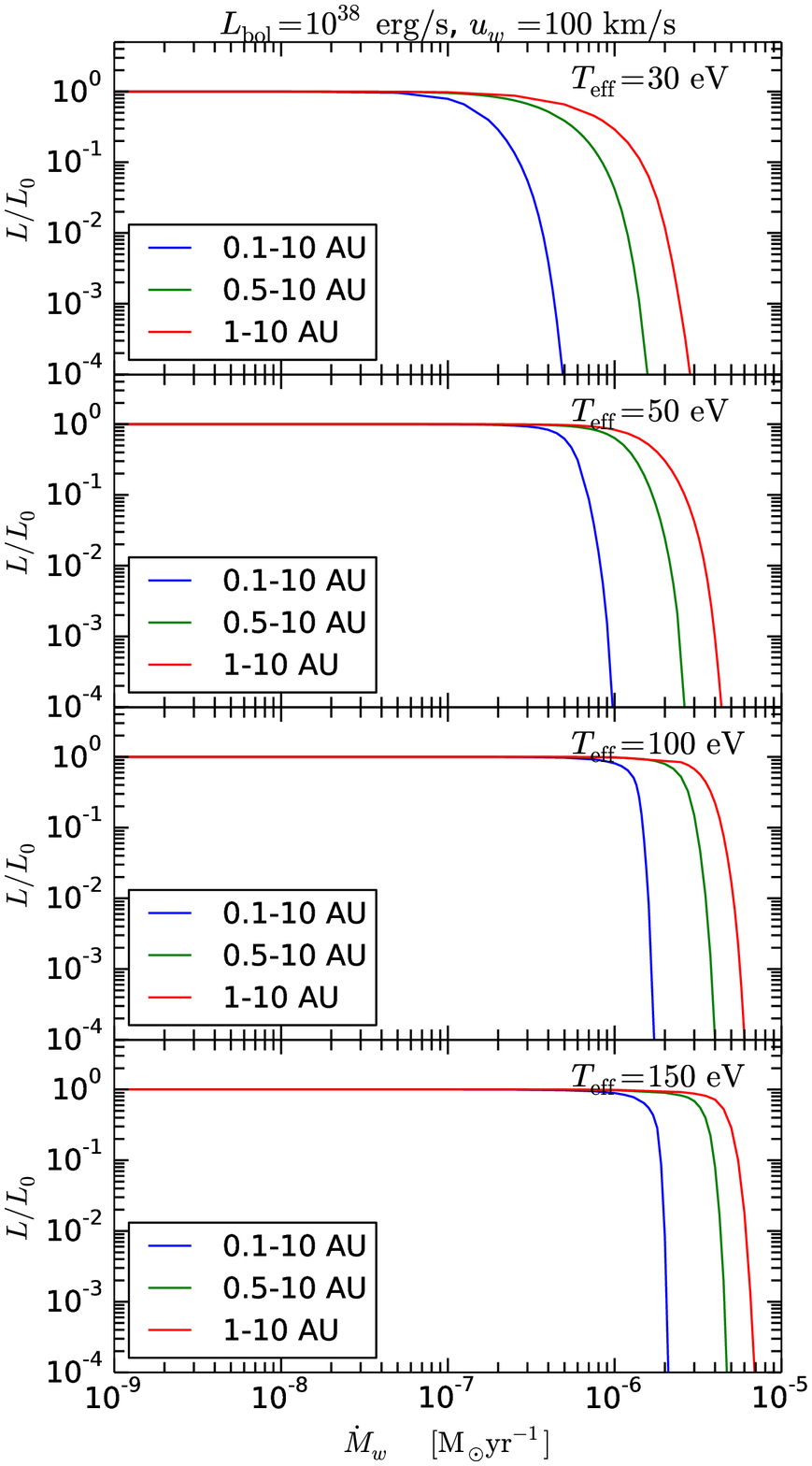}
    }    
    \caption{Attenuation of a super-soft source as a function of the mass outflow rate $\dot{M}_w$. The super-soft radiation is assumed to have a blackbody spectrum with luminosity of $L_{\mathrm{bol}}=10^{38}$ erg/s, and the luminosity emerging from the system is integrated from 300 eV to 1500 eV. The three columns of panels show results for different outflow velocities:  $u_w=10$ km/s (left) $u_w=50$ km/s (middle) and $u_w=100$ km/s (right). Different panels in each column  correspond to different values of the source temperature (marked in the upper-right corners of each panel) and different curves in each panel correspond to different values of the inner  radius of the gas shell, as marked in the plots. The outer boundary of the  shell is $r_{\mathrm{out}}=10$ AU in all cases.}
\label{Fig:atten_comb}
\end{figure*}

The behaviour shown in Figure \ref{Fig:absspectr_combined_1-10AU} is qualitatively similar for other configurations of $kT_{\mathrm{eff}}$, $r_{\mathrm{inn}}$, $r_{\mathrm{out}}$ and $u_w$, although the critical mass-loss rate where obscuration becomes significant is of course different for different combinations of parameters. Results of our calculations are summarised in  Figure \ref{Fig:atten_comb} where we show the attenuation curves as a function of the mass loss rate  for different values of the source temperature  and gas shell parameters. Table \ref{Table:NH_required_for_0.01_transmission} gives the $\dot{M}_{w,\mathrm{crit}}$  values required to produce an attenuation of a factor of  $10^2$. It also lists the corresponding column densities of the gas shell $N_H$.

\begin{table}
 \caption{Values of the mass-loss rate  and corresponding $N_H$, needed for an attenuation of $L_{\mathrm{obs}}/L_0 = 0.01$.}
 \centering
  \begin{tabular}{@{} c c c c c @{}}
  \hline
		$u_w$	& $T_{\mathrm{eff}}$	& $r_{\mathrm{inn}}, r_{\mathrm{out}}$	& $\dot{M}_{w,\mathrm{crit}}$		& $N_H$ \\ 

		[km/s]	& [eV]			& [AU]		& [M$_{\odot}\mathrm{yr}^{-1}$]	& [cm$^{-2}$] \\
    \hline
    \hline
		10.0 	& 30.0 			& 0.1, 10 	& 3.6$\times 10^{-8}$ 	& 5.2$\times 10^{22}$ \\
		" 	& '' 			& 0.5, 10 	& 1.3$\times 10^{-7}$ 	& 3.2$\times 10^{22}$ \\
		" 	& '' 			& 1.0, 10 	& 2.0$\times 10^{-7}$ 	& 2.6$\times 10^{22}$ \\
		" 	& 50.0 	 		& 0.1, 10 	& 8.2$\times 10^{-8}$ 	& 1.2$\times 10^{23}$ \\
		" 	& '' 	 		& 0.5, 10 	& 2.1$\times 10^{-7}$ 	& 5.8$\times 10^{22}$ \\
		" 	& '' 	 		& 1.0, 10 	& 3.5$\times 10^{-7}$ 	& 4.5$\times 10^{22}$ \\
		" 	& 100.0  		& 0.1, 10 	& 1.6$\times 10^{-7}$ 	& 2.3$\times 10^{23}$ \\
		" 	& ''  			& 0.5, 10 	& 3.5$\times 10^{-7}$ 	& 9.5$\times 10^{22}$ \\
		" 	& ''  			& 1.0, 10 	& 5.1$\times 10^{-7}$ 	& 6.6$\times 10^{22}$ \\
		" 	& 150.0  		& 0.1, 10 	& 2.0$\times 10^{-7}$ 	& 2.8$\times 10^{23}$ \\
		" 	& ''	  		& 0.5, 10 	& 4.1$\times 10^{-7}$ 	& 1.2$\times 10^{23}$ \\
		" 	& ''	  		& 1.0, 10  	& 6.1$\times 10^{-7}$ 	& 7.9$\times 10^{22}$ \\
    \hline
		50.0 	& 30.0 			& 0.1, 10 	& 1.8$\times 10^{-7}$ 	& 5.2$\times 10^{22}$ \\
		" 	& '' 			& 0.5, 10 	& 5.9$\times 10^{-7}$ 	& 3.2$\times 10^{22}$ \\
		" 	& '' 			& 1.0, 10 	& 1.0$\times 10^{-6}$ 	& 2.6$\times 10^{22}$ \\
		" 	& 50.0 			& 0.1, 10 	& 4.1$\times 10^{-7}$ 	& 1.2$\times 10^{23}$ \\
		" 	& '' 			& 0.5, 10 	& 1.1$\times 10^{-6}$ 	& 5.8$\times 10^{22}$ \\
		" 	& '' 			& 1.0, 10 	& 1.7$\times 10^{-6}$ 	& 4.5$\times 10^{22}$ \\
		" 	& 100.0 		& 0.1, 10 	& 8.0$\times 10^{-7}$ 	& 2.3$\times 10^{23}$ \\
		" 	& '' 			& 0.5, 10 	& 1.8$\times 10^{-6}$ 	& 9.5$\times 10^{22}$ \\
		" 	& ''	 		& 1.0, 10 	& 2.6$\times 10^{-6}$ 	& 6.6$\times 10^{22}$ \\
		" 	& 150.0 		& 0.1, 10 	& 9.9$\times 10^{-7}$ 	& 2.8$\times 10^{23}$ \\
		" 	& ''	 		& 0.5, 10 	& 2.2$\times 10^{-6}$ 	& 1.2$\times 10^{23}$ \\
		" 	& ''	 		& 1.0, 10 	& 3.1$\times 10^{-6}$ 	& 7.9$\times 10^{22}$ \\
    \hline
		100.0 	& 30.0 			& 0.1, 10 	& 3.7$\times 10^{-7}$ 	& 5.2$\times 10^{22}$ \\
		" 	& '' 			& 0.5, 10 	& 1.2$\times 10^{-6}$ 	& 3.2$\times 10^{22}$ \\
		" 	& '' 			& 1.0, 10 	& 2.0$\times 10^{-6}$ 	& 2.6$\times 10^{22}$ \\
		" 	& 50.0 			& 0.1, 10 	& 8.2$\times 10^{-7}$ 	& 1.2$\times 10^{23}$ \\
		" 	& '' 			& 0.5, 10 	& 2.2$\times 10^{-6}$ 	& 5.8$\times 10^{22}$ \\
		" 	& '' 			& 1.0, 10 	& 3.5$\times 10^{-6}$ 	& 4.5$\times 10^{22}$ \\
		" 	& 100.0 		& 0.1, 10 	& 1.6$\times 10^{-6}$ 	& 2.3$\times 10^{23}$ \\
		" 	& ''	 		& 0.5, 10 	& 3.5$\times 10^{-6}$ 	& 9.5$\times 10^{22}$ \\
		" 	& ''	 		& 1.0, 10 	& 5.1$\times 10^{-6}$ 	& 6.6$\times 10^{22}$ \\
		" 	& 150.0 		& 0.1, 10 	& 2.0$\times 10^{-6}$ 	& 2.8$\times 10^{23}$ \\
		" 	& ''	 		& 0.5, 10 	& 4.3$\times 10^{-6}$ 	& 1.2$\times 10^{23}$ \\
		" 	& ''	 		& 1.0, 10 	& 6.1$\times 10^{-6}$ 	& 7.9$\times 10^{22}$ \\
    \hline
\end{tabular} \label{Table:NH_required_for_0.01_transmission}
\end{table}

\subsection{Parameter dependence} \label{Subsect:Parameter.dependence}
\begin{figure}
    \centerline{\includegraphics[width=0.95\linewidth]{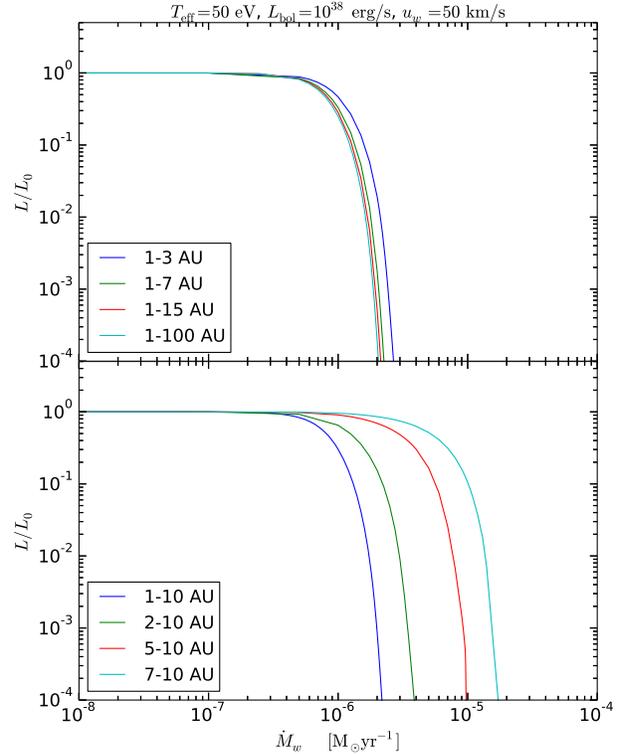}}
    \caption{Attenuation curves showing the effect of different values of $r_{\mathrm{out}}$ (top panel) and $r_{\mathrm{inn}}$ (bottom panel). In both panels, $kT_{\mathrm{eff}}=50$ eV, $L_{\mathrm{bol}}=10^{38}$ erg/s, $u_w=50$ km/s.}
    \label{Fig:atten_varying_inner_and_outer_radius}
\end{figure}

Figure \ref{Fig:atten_varying_inner_and_outer_radius} shows comparisons between attenuation curves for different combinations of inner and outer radii, as a function of the circumstellar mass loss rate. In agreement with the scaling suggested by Eq. (\ref{eq:mdot_crit}), the attenuation curve is moderately dependent on the value of the inner radius and nearly insensitive to the position of the outer boundary of the shell, as long as $r_{\mathrm{out}}/r_{\mathrm{inn}}$ is sufficiently large. 

\begin{figure}
    \centerline{\includegraphics[width=0.95\linewidth]{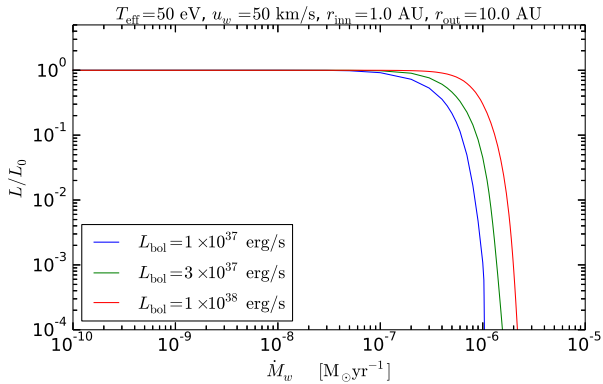}}
    \caption{Attenuation curves for different values of the luminosity of the central source. Other parameters are fixed at the following values:$kT_{\mathrm{eff}}=50$ eV, $u_w=50$ km/s.}
    \label{Fig:atten_l}
\end{figure}

Dependence of the attenuation curve on the luminosity of the central source is illustrated in Figure \ref{Fig:atten_l}. As one can see from the plot, it is somewhat weaker than predicted by Eq. (\ref{eq:mdot_crit}). This is because the derivation of Eq. (\ref{eq:mdot_crit}) did not take collisional ionisation into account.

As the gas density in the shell is determined by the ratio $\dot{M}_w/u_w$, the attenuation curves shift  along the $\dot{M}_w$ axis  linearly with   the outflow velocity. This is obvious from Figure \ref{Fig:atten_comb} and Table \ref{Table:NH_required_for_0.01_transmission} (cf. Eq. (\ref{eq:mdot_crit})).

\subsection{Chemical abundances} \label{Subsect:Dependence.on.chemical.abundances}
\begin{figure}
    \centerline{\includegraphics[width=0.95 \linewidth]{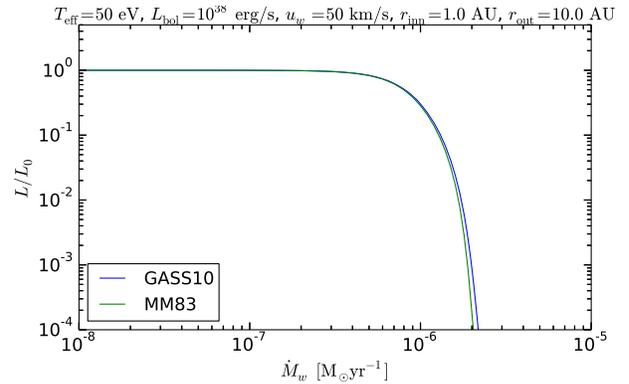}}
    \caption{Comparison between attenuation curves computed for the abundance table of \citep{Grevesse.et.al.2010} (marked \texttt{GASS10}) and the older one used in \citet{Nielsen.et.al.2013} (marked \texttt{MM83}), for $kT_{\mathrm{eff}}=50$ eV, $L_{\mathrm{bol}}=10^{38}$ erg/s, $r_{\mathrm{inn}}=1$ AU, $r_{\mathrm{out}}=10$ AU and $u_w=50$ km/s.}
    \label{Fig:atten_comb_50eV_GASS10vsMM83}
\end{figure}

The chemical abundances used in \citet{Nielsen.et.al.2013} were taken from \citet{Anders.Ebihara.1982} (for all other elements than helium) and \citet{Peimbert.Torres-Peimbert.1977} (for helium), as tabulated in \citet{Morrison.McCammon.1983}. The present study used the more updated chemical abundances table from \citet{Grevesse.et.al.2010}. Figure \ref{Fig:atten_comb_50eV_GASS10vsMM83} shows a comparison between the attenuation curves for the two abundance tables, when calculated with Cloudy using otherwise identical input parameters. For reference, the 'GASS10' curve in Figure \ref{Fig:atten_comb_50eV_GASS10vsMM83} is the same as the '1.0-10 AU' curve in the second panel from the top, middle column on Figure \ref{Fig:atten_comb}. As Figure \ref{Fig:atten_comb_50eV_GASS10vsMM83} shows, the difference between the transmission curves computed for the two abundance tables is insignificant. For comparison, a transmission ratio of $10^{-4}$ corresponds to a mass-loss rate of $\sim2.04\times10^{-6}$ M$_{\odot}\mathrm{yr}^{-1}$ for the MM83 abundances and $\sim2.18\times10^{-6}$ M$_{\odot}\mathrm{yr}^{-1}$ for GASS10.

%
\section{Discussion} \label{Sect:Discussion}
\begin{figure}
    \centerline{\includegraphics[width=0.95\linewidth]{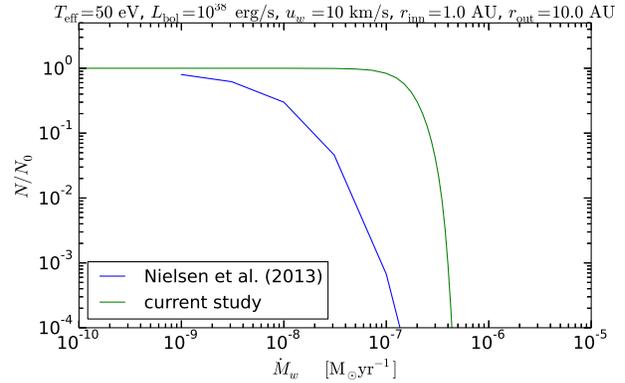}}
    \caption{Comparison between attenuation of number of ionising photons ($N/N_0$) in \citet{Nielsen.et.al.2013} and the current study, for $kT_{\mathrm{eff}}=50$ eV, $L_{\mathrm{bol}}=10^{38}$ erg/s, $r_{\mathrm{inn}}=1$ AU, $r_{\mathrm{out}}=10$ AU, and $u_w=10$ km/s.}
    \label{Fig:atten_compar_w_Nielsen_et_al_2013_Table_1}
\end{figure}

The results presented above demonstrate the importance of the proper account for photo-ionisation when considering attenuation of super-soft X-ray sources by circumbinary material.  In particular, the values of $\dot{M}_w$ required for significant obscuration of a canonical super-soft X-ray source are about an order of magnitude larger than those reported in \citet{Nielsen.et.al.2013}. Figure \ref{Fig:atten_compar_w_Nielsen_et_al_2013_Table_1} shows a direct comparison between the two works. In the figure, we plot the attenuation of the integrated number of photons emitted by an obscured source as given in Table 1 of \citet{Nielsen.et.al.2013} and those found in the current study for the same input parameters. As already explained above, the difference in these two curves stems from the detailed treatment of photo-ionisation of the circumbinary material by the radiation from a super-soft source.

The significantly larger mass-loss rates required for  obscuration of the super-soft source provide a stronger constraint on the possibility of nuclear burning massive white dwarfs being hidden by absorption by local circumstellar material. For example, in \citet{Nielsen.et.al.2013} it was suggested that for a 1 and 10 AU inner and outer radius, and a wind velocity of 10 km/s it would be possible for a massive, steady-burning white dwarf accreting at a few times $10^{-7}$ M$_{\odot}\mathrm{yr}^{-1}$ to be more or less fully obscured if $\lesssim$10\% of the material lost from the donor star did not become accreted onto the white dwarf but instead were available for obscuration in the circumstellar region. In the current study, we find for the same input values of parameters, that this amount of obscuration would require a mass loss into the circumstellar region of $\gtrsim50$\% of the mass-transfer rate, i.e. the amount of material lost into the circumstellar region would have to be of the order as the material being accreted onto the white dwarf. For wind velocities of $\sim 50-100$ km/s the circumstellar  mass loss rate should  exceed $\gtrsim90$\% of the mass-transfer rate, i.e. $\sim 10$ times more material should be lost into the circumstellar space than be accreted on to the white dwarf. Recalling that the steady-burning rate for a massive white dwarf is a few times $10^{-7}$ M$_{\odot}\mathrm{yr}^{-1}$, this means that to sustain both the steady burning super-soft X-ray source and the obscuring envelope the donor would have to lose $\gtrsim10^{-6}$ M$_{\odot}\mathrm{yr}^{-1}$ for a $\gtrsim 10^6$ years. The high mass-loss rate and the total mass lost (around a Solar mass or more) constrains the population of possible donors, and hence binary systems capable of producing significantly obscured, super-soft X-ray sources that eventually reach the Chandrasekhar mass and becomes type Ia supernovae.

\begin{figure*}
    \centerline{
    \includegraphics[width=1.0\linewidth]{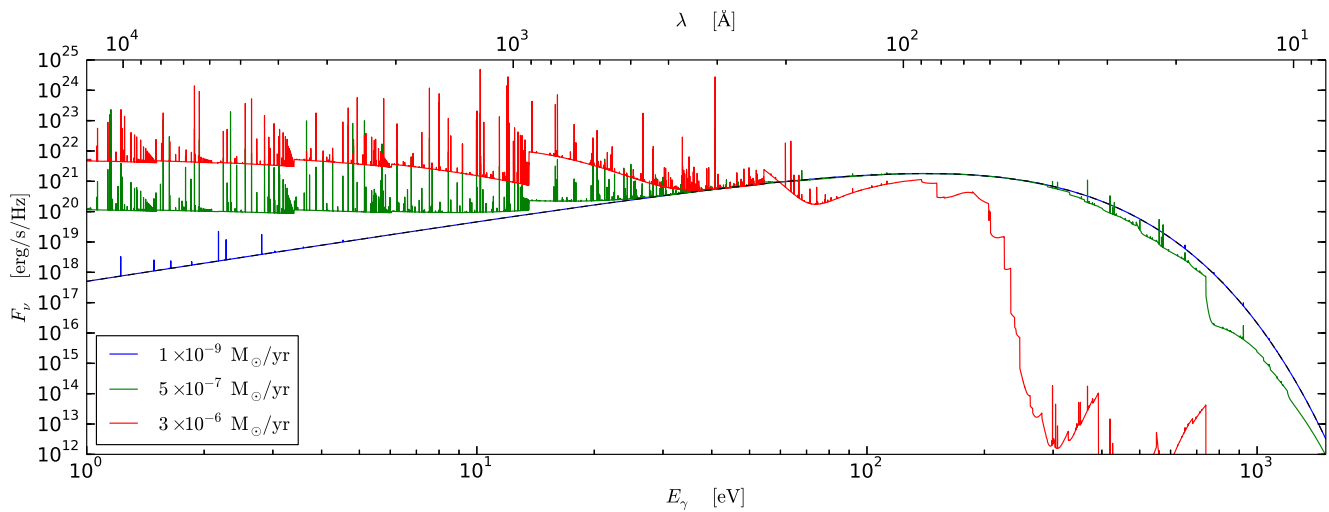}
    }    
    \caption{Broad band (infrared to X-ray) spectra of an attenuated super-soft X-ray source for three different attenuation levels:  an unobscured source ($\dot{M}=10^{-9}$ M$_{\odot}$/yr), moderately obscured source ($\dot{M}=5\times10^{-7}$ M$_{\odot}$/yr) and heavily obscured source ($\dot{M}=3\times10^{-6}$ M$_{\odot}$/yr).  The model parameters are at their default values: $kT_{\mathrm{eff}}=50$ eV, $L_{\mathrm{bol}}=10^{38}$ erg/s, $r_{\mathrm{inn}}, r_{\mathrm{out}}=$1 AU,10 AU. The dashed black line is the incident spectrum.}
\label{Fig:absspectr_combined_1-10AU_long_wlengths}
\end{figure*} 

Although we considered here a specific case of the circumbinary wind, interpreted broadly, our results show that a canonical super-soft source is capable of ionising about $N_H\sim 10^{22} - 10^{23}$ cm$^{-2}$ of material located at the distance of a $\sim$ few astronomical units from the source (Table \ref{Table:NH_required_for_0.01_transmission}).

Looking from a different perspective, a narrow but dense shell of gas with the hydrogen column density in excess of  $\sim 10^{23}$ cm$^{-2}$  around a single-degenerate type Ia supernova progenitor could  conceivably obscure the super-soft X-ray emissions from such a system.  Because of the large velocities of supernova ejecta (upwards of $\sim$ 10,000 km/s -- or $\sim$ 5.8 AU/day -- for type Ia supernovae) such a shell would be overtaken by the ejecta within a matter of days. Hence, upper limit measurements from e.g. radio observations would need to be made almost immediately after the supernova explosion to provide useful constraints on the amount of circumbinary material present at the $\sim$AU distances from the supernova. The example of nearby SN2011fe is probably to be considered a best case in this respect, as it was discovered and subjected to follow-up observations almost immediately after exploding \citep{Nugent.et.al.2011}. Even so, the radio limits placed on SN2011fe were based on observations conducted a little over two days after the supernova \citep{Chomiuk.et.al.2012}, which means that the blast wave would have moved to a distance of $\sim 12$ AU (assuming the above-mentioned ejecta velocity) at the time observations began. Hence, a very dense shell of material of, say, between 1 and 10 AU from the exploding object may conceivably have eluded detection, as the supernova ejecta  would have passed through the obscuring medium before observations began. More distant type Ia supernovae are typically discovered several days to weeks after the explosion, and in such cases radio observations should not be expected to yield upper limits capable of ruling out narrow configurations of circumstellar material. However, since single-degenerate progenitors of type Ia supernovae need to be super-soft X-ray sources for extended periods of time and require continuous mass-loss rates $\gtrsim10^{-6}\mathrm{M}\mathrm{yr}^{-1}$ to remain hidden, the extent and mass of the circumstellar gas surrounding such sources could be much larger than the narrow configurations mentioned above  and should therefore, theoretically at least, be detectable via post-supernova radio observations. Also, such extended systems should be detectable in pre-supernova observations in the recombination lines of hydrogen and helium and forbidden lines of metals  \citep[cf.][]{Woods.Gilfanov.2013}.

\subsection{Observational appearance of obscured SSS} \label{Subsect:Emission.lines}
Although  circumbinary obscuration is unlikely to explain the paucity of  super-soft sources in the context of progenitors of type Ia supernovae, it is plausible that stably nuclear-burning white dwarfs experience various degree of circumbinary obscuration. In particular, this may be one of the factors (along with the reasons intrinsic to  the population synthesis calculations) explaining the discrepancy between the number of super-soft sources predicted in population synthesis studies and their observed numbers (e.g. \citealt{Chen.et.al.2014,Chen.et.al.2015}). While quenching the bulk of the X-ray emission from the stably nuclear-burning white dwarf, circumbinary obscuration leads to the appearance of various emission signatures  at all wavelengths (including X-ray band itself), which can be exploited to search and identify such obscured super-soft X-ray sources.  

Figure \ref{Fig:absspectr_combined_1-10AU_long_wlengths} shows the output spectrum of an attenuated super-soft X-ray source in the broad wavelength range  from infrared to X-ray wavelengths. Spectra for three values of the mass-loss rate $\dot{M}_w$ are shown, corresponding to different obscuration levels in the X-ray band -- from completely unobscured to completely obscured source (cf. Fig.\ref{Fig:atten_comb}). In the completely unobscured case, the gas in the wind  is nearly fully  ionised and optically thin, and the attenuation at these mass-loss rates is negligible (save for a short period when the source initially 'turns on' and ionises the cloud). Such sources would be similar to naked super-soft sources when observed in X-rays, as discussed previously.  In the obscured case, the soft X-ray radiation is reprocessed to the optical and UV bands, giving rise to copious emission lines.  In the intermediate case of moderate obscuration a number of emission features can be observable in the X-ray band as well. 

\begin{figure*}
    \centerline{\includegraphics[width=0.65\linewidth]{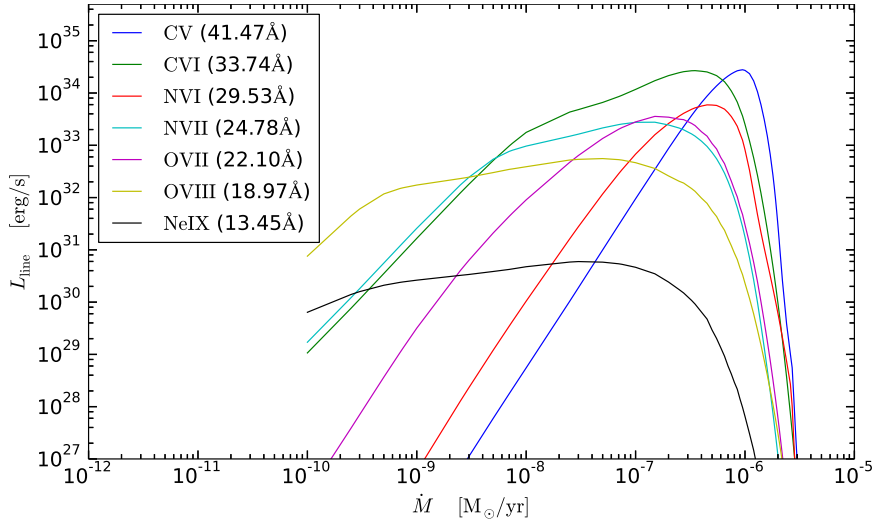}}
    \caption{Luminosity of the most prominent  X-ray emission lines vs the mass-loss rate, for our default model setup ($kT_{\mathrm{eff}}=50$ eV, $L_{\mathrm{bol}}=10^{38}$ erg/s, $u_w=50$ km/s, $r_{\mathrm{inn}}, r_{\mathrm{out}}=$1 AU,10 AU).}
    \label{Fig:Xray_emission_line_luminosities_vs_Mdot}
\end{figure*}

\subsubsection{X-ray lines} \label{Subsubsect:X-ray.features}
X-ray lines, although not the brightest among the emission features in our calculations, can carry away up to $\sim 10^{-4}-10^{-3}$ of the luminosity of the central source. The brightest among them are resonant lines of H-like ions and He-like triplets of the low-Z metals. In Figure \ref{Fig:Xray_emission_line_luminosities_vs_Mdot} we show luminosities of several of the brightest X-ray lines as a function of the mass-loss rate.  As the amount of the circumbinary material increases with the mass-loss rate, luminosities of these lines initially grow with $\dot{M}_w$, until obscuration in the X-ray region sets in  at $\sim 10^{-6}$ M$_{\odot}$/yr (see Figure \ref{Fig:atten_comb}). At this point, the X-ray line luminosities plummet due to absorption by the neutral outer layers of the wind. The behaviour of particular lines is mainly determined  by the changes of relative abundances of respective ions, with the intrinsic line ratios being characteristic for photo-ionised plasma with ionisation parameter of $\xi\approx 37.4\, L_{38}\, \dot{M}^{-1}_{-6}\, u_{100}$.  However, as the mass-loss rate approaches the critical value, the escaping line luminosities and their ratios are significantly modified  by the absorption by the neutral outer part of the wind, giving rise to the patterns shown in Fig.\ref{Fig:Xray_emission_line_luminosities_vs_Mdot}.

\cite{Ness.et.al.2013} studied high-resolution spectra for a number of super-soft X-ray sources in local galaxies. Based on these observations, they suggested a subdivision of super-soft X-ray sources into two classes: SSa ('absorbed') and SSe ('emitting') -- the former having generally blackbody-like spectra with some prominent absorption features, and the latter showing strong emission features (comparable in strength to the continuum) superimposed on weak blackbody spectra. When correlated with the inclinations of the systems of that study (and with the caveat that inclination determination is quite problematic in most of the cases) members of the SSa class were predominantly high-inclination systems, while SSe systems were predominantly viewed more edge-on. This led Ness and collaborators to suggest that the spectral differences of the two classes might arise as a result of viewing-angle, rather than from intrinsic differences between the systems themselves, conceptually similar to the viewing-angle effect in the unified model for AGNs. Although the direct comparison with observed spectra is outside the scope of the present work, our study does recreate several of the features of the SSe class reported in \cite{Ness.et.al.2013} in the interval where the central sources transition from unobscured to fully obscured regime. The two strongest emission features in the spectra of U Sco and CAL 87 on Figure 3 in \cite{Ness.et.al.2013} correspond to the O VIII-doublet at 18.97\AA , 18.98\AA , and the line blend consisting of the N VII-doublet at 24.78\AA , 24.79\AA , and the N VI at 24.89\AA . We find the same features are prominent in our spectra, as seen on our Figure \ref{Fig:absspectr_combined_1-10AU}. Several other, less pronounced lines in our spectra can be also found in the observed  spectra of U Sco and CAL 87, e.g. Ne IX at 13.45\AA , O VIII at 21.6\AA , and possibly three N VI lines around 29\AA\ (28.78\AA , 29.1\AA , 29.5\AA). At lower X-ray energies the similarities between our spectra and those of \cite{Ness.et.al.2013} become less obvious, presumably due to interstellar absorption in the observed spectra. The fact that our relatively simple model setup can reproduce some of the features of the SSe spectra disucssed in \cite{Ness.et.al.2013} is not in itself surprising, since any model with significant incident luminosity in X-rays which interacts with a Solar metallicity gas is likely to produce a somewhat similar set of emission features. However the luminosities of these lines, their ratios and the presence and strength of absorption lines and edges are determined by the details of the  model such as geometry of the gas and emission pattern, ionisation parameter and  chemical abundances. Detailed analysis of these effects is beyond the scope of the present paper.

\subsubsection{Optical lines} \label{Subsubsect:Lower.energy.features}

As the mass-loss rate $\dot{M}_w$ grows towards full obscuration, the X-ray emission is gradually quenched, and the absorbed energy increasingly re-emitted at UV and optical energies. In this regime, copious emission lines in the UV, optical and infrared bands are produced in the photo-ionised wind material. The majority of these lines (except for extreme UV lines), unlike X-ray lines, escape the wind material without significant attenuation and therefore may serve as a valuable diagnostic tool in the search for and identification absorbed super-soft sources.

\begin{figure}
    \centerline{\includegraphics[width=0.95\linewidth]{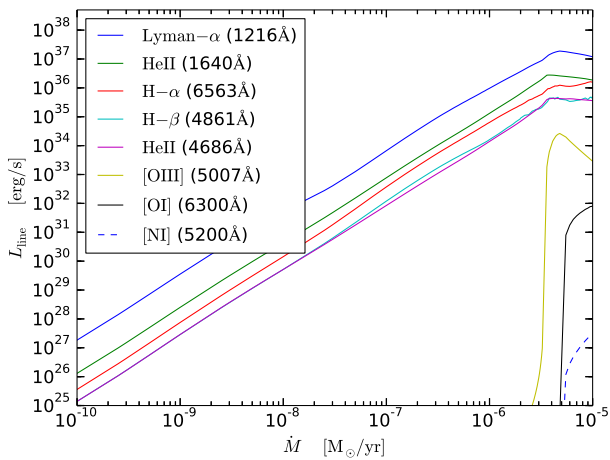}}
    \caption{Luminosity of several most prominent  optical and UV lines vs the mass-loss rate for our default model  ($kT_{\mathrm{eff}}=50$ eV, $L_{\mathrm{bol}}=10^{38}$ erg/s, $u_w=50$ km/s, $r_{\mathrm{inn}}, r_{\mathrm{out}}=$1 AU,10 AU). The 'wavy' behaviour of the H$_{\alpha}$ and H$_\beta$ curves at $\dot{M}_w \gtrsim 10^{-6}$ M$_{\odot}$/yr is a numerical effect, arising from optical depth convergence problems in Cloudy at large optical depths. This is a known problem that will presumably be addressed in an upcoming update of the code (Cloudy Collaboration, private communication).}
    \label{Fig:emission_line_luminosities_vs_Mdot}
\end{figure}

Figure \ref{Fig:emission_line_luminosities_vs_Mdot} shows the luminosity of a number of observationally relevant lines of hydrogen and helium as well as forbidden lines of metals  as a function of the mass-loss rate. The luminosity of hydrogen and helium recombination lines grows monotonically with mass-loss rate and then saturate as the central source becomes completely obscured (at roughly $\dot{M}_w\sim {\rm few}\times 10^{-6}$ M$_{\odot}$/yr in our default model, see Figure \ref{Fig:atten_comb}). At this point, the ionised medium becomes confined within the neutral outer layers of the wind, and forbidden lines begin to appear as well, produced at the interface of the ionised and neutral gas. At $\dot{M}_w \gtrsim 3 - 5 \times10^{-6}$ M$_{\odot}$/yr, the Ly$_\alpha$ luminosity of the gas cloud reaches $\sim 10\%$ of the total luminosity of the central source.

\begin{figure}
    \centerline{\includegraphics[width=0.95\linewidth]{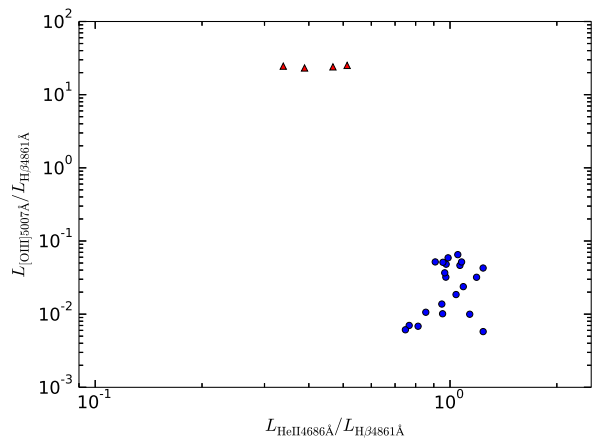}}
	\caption{Comparison between the  [OIII] 5007\AA\  and HeII 4686\AA\ to H$_\beta$ line ratios expected from the ionised nebulae in the low density ISM  (Rappaport et al. (1994), red triangles) and the corresponding ratios expected for an obscured  super-soft source imbedded in the dense wind considered in this work (blue circles). The wind case was computed  for our default model ($kT_{\mathrm{eff}}=50$ eV, $L=10^{38}$ erg/s, $r_{\mathrm{inn}}, r_{\mathrm{out}}=1.0,10.0$ AU); the points shown are for $\dot{M}_w$ values for which the [OIII] luminosity is non-negligible, i.e. $\dot{M}_w \gtrsim 3.5\times10^{-6}$ M$_{\odot}$/yr.}
    \label{Fig:RCKM_compar_plot}
\end{figure}

\cite{Rappaport.et.al.1994} modelled the effect of persistent, unobscured, super-soft X-ray sources on the surrounding (low density) interstellar medium. Their results showed that such sources should develop parsec-sized ionisation nebulae which should exhibit strong emission in forbidden optical lines, most importantly [O III] 5008{\AA}, as well as in recombination lines of hydrogen and helium, including He II 4686{\AA}. They suggested using forbidden oxygen lines ([OI] and [OIII]) as a diagnostic tool to find ionised ISM nebulae around on unobscured super-soft X-ray sources \citep{Rappaport.et.al.1994,Di.Stefano.et.al.1995}. 
Much higher gas densities  in the wind  ($\sim 10^9$ cm$^{-3}$ in the wind vs. $\sim 1$ cm$^{-3}$ in the ISM) make forbidden lines much weaker in our case. 
Figure \ref{Fig:RCKM_compar_plot} shows a comparison of the [OIII] 5007\AA\ and HeII 4686\AA\  to H$_\beta$ line ratios in the case of the obscuration by the wind with those expected from the ionisation nebulae computed by \cite{Rappaport.et.al.1994} (their Figure 5). As one can see from the plot, a super-soft source  obscured by the dense wind  considered in this work will be easily  distinguishable from the case of the ionised nebula around a super-soft source imbedded in the low density ISM.\footnote{A caveat here is that such a nebula has been so far detected only around one super-soft source \citep{Remillard.et.al.1995,Woods.Gilfanov.2015}.}

To conclude, the  observational signatures of a fully obscured super-soft X-ray sources should be significant (up to $\sim10^{37}$ erg/s for a $10^{38}$ erg/s source) Ly$_\alpha$ emission, coupled with the low  [OIII]-to-H$_\beta$ ratio, of $\sim 10^{-2}$, and relatively high HeII 4686\AA-to-H$_\beta$ ratio,  $\sim 1$. These characteristics make obscured super-soft sources an observationally distinct class of compact line emission sources. 

\subsection{The inner wind radius} \label{Subsect:Caveats}
The critical mass-loss rate at which obscuration of the central source becomes significant shows moderate dependence on the assumed inner radius of the wind. Eq. \ref{eq:mdot_crit_H} predicts that $\dot{M}_{\rm w,crit}\propto r_{inn}^{1/2}$, whereas numerical calculations with Cloudy show marginally stronger scaling. The moderate $r_{\rm inn}$-dependence  is a consequence of the two assumptions made in our calculations: (i) that the $\rho_{gas}\propto r^{-2}$ density profile extends to infinitely small radii and (ii) that the source of radiation is located in the locus of this density distribution, i.e. at the point $r=0$. These two assumptions combined together lead to the strong dependence of the total recombination rate in the gas shell on its inner radius, eq.(\ref{eq:rate_i}). To eliminate these assumptions, one would need to specify the driving mechanisms behind the mass loss and accurately consider the physical conditions at the inner wind boundary, which is far beyond the scope of this work. Below we consider the  example of symbiotic systems to attribute some physical meaning to the inner radius in our simulations. 

Recent modelling of the red giant winds \citep{Suzuki.2007} demonstrated that the  $\rho_{gas}\propto r^{-2}$ density profile may indeed be a valid representation  of the wind structure down   to rather small radii, as low as  $\sim$0.1 AU. The source of ionising radiation (i.e. the white dwarf), however, is offset by the distance $a$ (binary separation) from the singular point of the gas distribution. In the absence of the singularity, the dependence of the total recombination rate in the gas shell on its inner radius is insignificant (as long as $r_{\rm inn}\ll r_{\rm out}$) and the only role of the $r_{\rm inn}$  in this case is to provide the normalisation of the density distribution, via the mass continuity equation $\dot{M}_w=4\pi\,r^2\,\rho(r)\, u_w$. Therefore it is natural to set $r_{\rm inn}=a$ in this case, i.e. the inner radius is of the order of the orbital separation in the binary system. Note also, that  the 3-dimensional nature of the problem becomes essential at $r\sim a$, as the red giant will be casting a shadow onto the wind material located behind it. Obviously, the simplified 1-dimensional consideration in this paper is applicable only to the lines of sight which are not passing close to the donor star.

To conclude, while the picture of a spherical shell confined between the inner and outer boundaries is a simplification of the real geometry of a circumbinary gas cloud, our model is a useful first approximation. This is the geometry considered in the previous work \citep{Nielsen.et.al.2013} and its use here facilitates comparison with those results and helps to isolate the effect of photo-ionisation. Future work (currently in progress) will aim at dealing with more realistic density structures and source spectra than those discussed here.

%
\section{Conclusions} \label{Sect:Conclusion}

We have presented full photo-ionisation simulations of the interaction between the radiation from a super-soft X-ray source and a circumstellar cloud of Solar metallicity gas. This seeks to model a single-degenerate  system that has lost material from either the donor or the accretor into the circumstellar region. The aim has been to determine the amount of material (parametrised by a circumstellar mass-loss rate) required to significantly obscure the X-ray emission from the accretor.

To simulate the effect of the X-ray source on the gas cloud we used the one-dimensional photo-ionisation code Cloudy \citep{Ferland.et.al.2013}. Our analysis explored intervals of source temperature, luminosity, inner and outer radii of the gas cloud, and wind velocity, to cover the range of values for realistic systems.

We found that the amount of circumstellar matter required to hide a super-soft X-ray source from our current observational capabilities is about an order of magnitude larger than that found in \citet{Nielsen.et.al.2013}. In particular, the required mass-loss rates from the binary typically exceeds $10^{-7}$ to a few times $10^{-6}$ M$_\odot \mathrm{yr}^{-1}$ for the parameters considered  in the 'default' case in this work. Since the donor star has to provide both the material being accreted onto the white dwarf (and hence powering its X-ray emission) and the circumstellar material, this places significant constraints on the mass of the donor star.

We discussed the observational appearance of hypothetical nuclear burning white dwarfs obscured by a wind. Such systems would appear as luminous ($\sim 10^{37}$ erg/s) compact sources of Ly$_\alpha$ emission with strong HeII recombination lines  (HeII 4686\AA\ to H$_\beta$ ratio  $\sim 1$ and  suppressed forbidden lines of metals, e.g. [OIII]-to-H$_\beta$ ratio, of $\sim 10^{-2}$), making them distinct from photo-ionised nebulae predicted to exist around super-soft X-ray sources imbedded in the low density ISM.

%
\section{Acknowledgements}
The authors thank the anonymous referee for providing constructive and inspiring comments that have helped to improve the original version of the manuscript.
MG acknowledges partial support by Russian Scientific Foundation (RNF), project 14-22-00271.

%

\label{lastpage}

\end{document}